\documentclass[twocolumn,showpacs,preprintnumbers,nofootinbib,amsmath,amssymb]{revtex4}

\usepackage{euscript,amssymb,amsmath}

\usepackage{graphicx}
\usepackage{epsfig}

\newcommand{\be}[1]{\begin{equation}\label{#1}}
\newcommand{\ee}{\end{equation}}
\newcommand{\ba}[1]{\begin{eqnarray}\label{#1}}
\newcommand{\ea}{\end{eqnarray}}
\newcommand{\rf}[1]{(\ref{#1})}
\newcommand{\nn}{\nonumber}

\newcommand{\rt}{r_3}

\begin{document}

\title{Latent solitons, black strings, black branes, \\
and equations of state in Kaluza-Klein models}

\author{Maxim Eingorn\dag}\email{maxim.eingorn@gmail.com}
\author{Orival R. de Medeiros\ddag}\email{orival@ufpa.br}
\author{Lu\'{i}s C. B. Crispino\ddag}\email{crispino@ufpa.br}
\author{Alexander Zhuk\dag\ddag}\email{ai_zhuk2@rambler.ru}

\affiliation{\dag Astronomical Observatory and Department of Theoretical Physics,
 Odessa National University, Street Dvoryanskaya 2, Odessa 65082, Ukraine\\}

\affiliation{\ddag Faculdade de F\'{i}sica, Universidade Federal do Par\'{a}, 66075-110,
Bel\'{e}m, PA, Brazil}

%\date{}

\begin{abstract}
In Kaluza-Klein models with an arbitrary number of toroidal internal spaces, we
investigate soliton solutions which describe the gravitational field of a massive compact
object. Each $d_i$-dimensional torus has its own scale factor $C_i,\; i=1,\ldots ,N$,
which is characterized by a parameter $\gamma_i$. We single out the physically interesting
solution corresponding to a point-like mass. For the general solution we obtain equations
of state in the external and internal spaces. These equations demonstrate that the
point-like mass soliton has dust-like equations of state in all spaces. We also obtain the
parameterized post-Newtonian parameters, which give the possibility to obtain the
formulas for perihelion shift, deflection of light and time delay of radar echoes.
Additionally, the gravitational experiments lead to a strong restriction on the
parameters of the model: $\tau = \sum_{i=1}^N d_i\gamma_i= -(2.1\pm 2.3)\times 10^{-5}$.
The point-like mass solution with $\gamma_1=\ldots=\gamma_N=(1+\sum_{i=1}^Nd_i)^{-1}$
contradicts this restriction. The condition $\tau=0$ satisfies the experimental limitation
and defines a new class of solutions which are indistinguishable from general relativity.
We call such solutions latent solitons. Black strings and black branes with $\gamma_i=0$
belong to this class. Moreover, the condition of stability of the internal spaces singles
out black strings/branes from the latent solitons and leads uniquely to the black string/brane
equations of state $p_i=-\varepsilon/2,\, i=1,\ldots ,N$, in the internal spaces and to
the number of the external dimensions $d_0=3$. The investigation of multidimensional static
spherically symmetric perfect fluid with dust-like equation of state in the external
space confirms the above results.
\end{abstract}


\maketitle

\vspace{.5cm}

%%%%%%%%%%%%%%%%%%%%%%%%%%%%%%%%%%%%%%%%%%%%%%%%%%%%%%%%%%%%%%%%%

\section{\label{sec:1}Introduction}

\setcounter{equation}{0}

Modern observational phenomena, such as dark energy and dark matter, are the
great challenge  for present cosmology, astrophysics and theoretical physics. Within the
scope of the standard models, it has still not been offered a satisfactory explanation
to these problems. This forces the search of solutions to these problems beyond
the standard models, for example, by considering models with
extra dimensions. This generalization follows from the modern theories of
unification such as superstrings, supergravity, and M-theory, which have the most
self-consistent formulation in spacetimes with extra dimensions. Obviously, these
physical theories should be consistent with observations. In a previous paper
\cite{EZ4}, two of the present authors have examined the family of 5-dimensional
soliton solutions (see \cite{Kramer,soliton,Davidson}) which describes the gravitational
field of massive compact objects in spacetimes with compact toroidal extra dimensions.
Among these solutions, the one which corresponds to a point-like massive source has been singled out.
At a first glance, this is a good physical approximation for astrophysical objects
in the weak-field limit because it works very well in general relativity. However,
it has been found that such approach contradicts famous gravitational experiments (perihelion
shift,  light deflection, and time delay of radar echoes) in 5-dimensional
spacetime \cite{EZ4,EZ3}. The only compact astrophysical objects that satisfy
the observational data with the same accuracy as general relativity, are objects
with a dust-like equation of state ($p_0=0$) in our three dimensions and the equation
of state $p_1=-\varepsilon/2$ in the fifth dimension. Black strings have
such equations of state. Additionally, it was shown that these equations of state
satisfy the necessary condition of the internal space stabilization (see also
\cite{Zhuk}). This is a very strong bound on the equations of state for the perfect fluid
in 5-dimensional Kaluza-Klein models. Therefore, it is important to understand how
common is this restriction. This was the main reason for our further investigations on the subject.
%The results of these studies are presented in this article.

Multidimensional soliton solutions were found in Refs. \cite{Ivashchuk,FIM,Leon1,Leon2}. In
the present paper, we investigate the solution from Ref. \cite{Leon1} because, to our
knowledge, it is the most general form of solitons in the case of an arbitrary number
$N$ of toroidal internal spaces. Our investigation shows that it is a very fruitful idea
to consider the most general case. In particular, we obtain the most general form of
the equations of state for a soliton matter source and we find an experimental restriction for the
parameters of the general solution. However, the most important advantage of such
approach is the discovery of a new class of solutions. These solutions satisfy the
well known gravitational experiments at the same level of accuracy as general relativity.
With these experiments only, it is impossible to differ these new solutions from
general relativity. For this reason we call them {\it latent solitons}. Black strings
and black branes belong to this class of solutions. We show that only black strings and
black branes have equations of state in the internal spaces which do not spoil the condition
of the internal space stabilization. All these conclusions are confirmed by a general
analysis of the multidimensional static spherically symmetric perfect fluid with dust-like
equation of state in the external (our) space. In the case of three-dimensional external
space, such perfect fluid describes observable astrophysical objects (e.g., the Sun)
in Kaluza-Klein models. Therefore, our investigation demonstrates that the condition of
stability of the internal spaces (i) singles out black strings/branes among the latent
solitons, (ii) leads uniquely to the black string/brane equations of state $p_i=-\varepsilon/2
\, ,\; i=1,\ldots ,N$, in the internal spaces, and (iii) to the number of the
external dimensions $d_0=3$.

The remainder of this paper is structured as follows. In Sec. II we consider a family
of the most general soliton solutions and single out one of these solutions which
corresponds to the physically important case of a point-like massive source. In Sec. III we
obtain equations of state for soliton matter sources in the most general case and show
that the point-like massive solution has dust-like equations of state in the external space as well as
in all internal spaces. In Sec. IV we find the parameterized post-Newtonian (PPN)
parameters in the general case. These parameters give us the possibility to derive the formulas
for perihelion shift, deflection of light and time delay of radar echoes. Using the
constraint which comes from the Cassini spacecraft experiment, we obtain in the general
case a strong restriction on the solitonic parameter $\tau=-(2.1\pm 2.3) \times 10^{-5}$.
The particular case $\tau=0$ defines a new class of latent solitons. We present the equations
of state for latent solitons as well as for black branes, which are a particular case
of latent solitons. In Sec. V, we investigate a multidimensional static spherically
symmetric perfect fluid with dust-like equation of state ($p_0=0$) in the external space
and arbitrary equations of state ($p_i=\omega_i\varepsilon$) in the internal spaces.
We demonstrate that the agreement with gravitational experiments leads to the latent soliton
equations of state in the internal spaces. The internal space stability condition singles
out the black brane equations of state. Our main results are summarized in Sec. VI.

%%%%%%%%%%%%%%%%%%%%%%%%%%%%%%%%%%%%%%%%%%%%%%%%%%%%%%%%%%%%%%%%%%%%%%%%%%%%%%%%%%%%%%%%%%%%%%%%
%%%%%%%%%%%%%%%%%%%%%%%%%%%%%%%%%%%%%%%%%%%%%%%%%%%%%%%%%%%%%%%%%%%%%%%%%%%%%%%%%%%%%%%%%%%%%%%%

%%%%%%%%%%%%%%%%%%%%%%%%%%%%%%%%%%%%%%%%%%%%%%%%%%%%%%%%%%%%%%%%%%%%%%%%%%%%%%%%%%
%%%%%%%%%%%%%%%%%%%%%%%%%%%%%%%%%%%%%%%%%%%%%%%%%%%%%%%%%%%%%%%%%%%%%%%%%%%%%%%%%%

\section{\label{sec:2}Soliton metrics}

\setcounter{equation}{0}

As we have already pointed out in the Introduction, the 5-dimensional soliton solutions
\cite{Kramer,soliton,Davidson} have been generalized to an arbitrary number of
dimensions in Refs. \cite{Ivashchuk,FIM,Leon1,Leon2}. To our knowledge, the most
general form of these solutions was given in Ref. \cite{Leon1}, and in isotropic
(with respect to our three-dimensional space) coordinates it reads
%%%%%
\ba{2.1}
&{}& ds^2 = A(r_3)c^2dt^2+B(r_3)(d\rt^2 +\rt^2d\Omega^2_2)\nn\\
&+&\sum\limits_{i=1}^{N}C_{i}(r_3)ds^2_{i}=\left(\frac{a\rt-1}{a\rt+1}\right)^{2\theta}
c^2dt^2\nn \\
&-&\left(1-\frac{1}{a^2\rt^2}\right)^2\left(\frac{a\rt+1}{a\rt-1}\right)^{2\theta(1-\tau)}
(d\rt^2 +\rt^2d\Omega^2_2)\nn\\
&-& \sum\limits_{i=1}^{N}\left(\frac{a\rt+1}{a\rt-1}\right)^{2\theta\gamma_{i}}ds^2_{i}
\, , \ea
%%%%
where $r_3$ is the length of the radius vector in three-dimensional space,
$ds^2_{i}=\sum_{j=1}^{d_i}d\xi^2_{(i)j}$ is the line element of the $d_i$-dimensional torus, and the
parameters $\tau$, $\theta $ and $\gamma_i$ satisfy
the condition~\cite{footnote1,footnote1a}
%%%%%%
\be{2.2} \theta^2[(\tau - 1)^2 +\sigma +1] = 2\, ,\quad \tau\equiv\sum\limits_{i=1}^{N}
d_{i}\gamma_{i}\, , \quad \sigma \equiv \sum\limits_{i=1}^{N}d_{i}\gamma^2_{i}\, .
\ee
%%%%%%

In the weak-field limit $1/(ar_3)\ll 1$, the metric coefficients are given by
%%%%%%
\ba{2.3}
A(r_3)&\approx& 1-\frac{4\theta}{ar_3}+\frac{16\theta^2}{a^2}\frac{1}{2 r_3^2}\, ,\\
B(r_3)&\approx&-1-\frac{4\theta(1-\tau)}{ar_3}\label{2.4}\, ,\\
C_i(r_3)&\approx&-1-\frac{4\theta \gamma_i}{ar_3}\label{2.5}\, .
\ea
%%%%%%
These expansions will help us to define important properties of the soliton solution
\rf{2.1}, for example, observational restrictions on the parameters of solitons and
equations of state for the matter source. These formulas are also useful to single
out the case of a point-like mass $m$ at rest as a matter source.

In the weak-field limit, the line element of a point-like mass $m$ at rest in a
$(1+D)$-dimensional spacetime with toroidal extra dimensions is \cite{EZ3}
%%%%%
\ba{2.6} &{}&ds^2\approx\left(1-\frac{r_g}{r_3}+\frac{r_g^2}{2r_3^2} \right)c^2dt^2\nn \\
&-&\left(1+\frac{1}{D-2}\, \frac{r_g}{r_3}\right)\left(dr_3^2+\rt^2d\Omega^2_2\right)\nn \\
&-&\left(1+\frac{1}{D-2}\, \frac{r_g}{r_3}\right)\sum\limits_{i=1}^{N}ds^2_{i}\, , \ea
%%%%%
where $r_g=2G_Nm/c^2$, with $G_N$ being the Newtonian gravitational constant.
The comparison of the metric coefficients \rf{2.3}-\rf{2.5} with the
corresponding metric coefficients in Eq. \rf{2.6} shows that for
the point-like mass, we have:

(i) The equality
%%%%%
\be{2.7}
\frac{4\theta}{a}=r_g
\ee
%%%%%
holds.
It follows that sign $a =$ sign $\theta$.
Because the solution \rf{2.1} is invariant under the simultaneous
change $a\to -a, \theta \to -\theta$, we can choose $a,\theta >0$.

(ii) The parameters $\gamma_i$ should take the same
value for all internal spaces, namely:
%%%%%%
\be{2.8} \gamma_1=\gamma_2=\ldots =\gamma_N =\frac{1}{1+D'}\, , \ee
%%%%%%
where $D'=\sum_{i=1}^{N}d_i = D-3$ is the total number of extra dimensions.

(iii) The parameters $\theta$ and $a$ are given by
%%%%
\be{2.9} \theta = \sqrt{\frac{2(1+D')}{2+D'}}\, , \quad a = \frac{4}{r_g}
\sqrt{\frac{2(1+D')}{2+D'}}\, , \ee
%%%%%%
where we also took into account the constraint \rf{2.2} and the relation \rf{2.7}.
Therefore, Eqs. \rf{2.7}-\rf{2.9} completely define the point-like mass soliton, i.e., the
solution where delta-shaped $T_{00}$ is the only non-zero component of the
energy-momentum tensor. To demonstrate it, in the next section we derive equations of
state for the general soliton solution \rf{2.1}.

%%%%%%%%%%%%%%%%%%%%%%%%%%%%%%%%%%%%%%%%%%%%%%%%%%%%%%%%%%%%%%%%%%%%%%%%%%%%%%%%%%%%%%%%%%%%%%%%
%%%%%%%%%%%%%%%%%%%%%%%%%%%%%%%%%%%%%%%%%%%%%%%%%%%%%%%%%%%%%%%%%%%%%%%%%%%%%%%%%%%%%%%%%%%%%%%%

\section{\label{sec:3}Equations of state. General case}

It is worth noting that the dependence of the metric coefficients only on $r_3$ in
Eq. \rf{2.1} means that the matter source for such metrics is uniformly
``smeared" over extra dimensions \cite{EZ1,EZ2}. It is clear that in this case
the non-relativistic gravitational potential depends only on $r_3$ and exactly
coincides with the Newtonian one. Since the function $A(r_3)$ is the metric
coefficient $g_{00}$, we obtain $4\theta /a=r_g=2G_Nm/c^2$, and  the expansions
\rf{2.3}-\rf{2.5} become
%%%%%
\ba{3.1}
A(r_3)&\approx& 1-\frac{r_g}{r_3}+\frac{1}{2}\frac{r_g^2}{r_3^2}\, ,\\
B(r_3)&\approx& -1-(1-\tau)\frac{r_g}{r_3}\label{3.2}\, ,\\
C_i(r_3)&\approx&-1-\gamma_i\frac{r_g}{r_3}\label{3.3}\, .\ea
%%%%%
From these expressions, we can easily get the perturbations
$h_{00}=-r_g/r_3, h_{\alpha\alpha}=-(1-\tau)r_g/r_3$ and
$h_{\mu_i\mu_i}=-\gamma_ir_g/r_3$, of the order of $1/c^2$ over the flat
spacetime, that gives us the possibility to find the components of Ricci
tensor up to the same order, namely:
%%%%%%%
\ba{3.4}
R_{00}&\approx& \frac{1}{2}\triangle h_{00}= \frac{1}{2}k_Nm
\delta({\bf r}_3)c^2=\frac{1}{2}k_N\rho_3 c^2\, ,\\
R_{\alpha\alpha}&\approx& \frac{1}{2}\triangle h_{\alpha\alpha}
=\frac{1}{2}(1-\tau)k_N\rho_3 c^2,\quad\alpha=1,2,3\, ,\label{3.5}\\
R_{\mu_i\mu_i}&\approx& \frac{1}{2}\triangle h_{\mu_i\mu_i}
=\frac{1}{2}\gamma_{i}k_N\rho_3 c^2\, ,\label{3.6}\\
\mbox{with}\nn\\
\mu_i &=& 1+ \sum_{j=0}^{i-1} d_j,\; \ldots \; , d_i + \sum_{j=0}^{i-1}
d_j \, ;\quad i=1,\ldots,N\nonumber\, ,
%&{}&\mu_i=4+\sum\limits_{j=1}^{i-1}d_i,\ldots,3+\sum\limits_{j=1}^id_i,\quad i=1,\ldots,N\nonumber\, .
\ea
%%%%%
where $d_0=3$, $k_N\equiv 8\pi G_N/c^4$ and $\triangle = \delta^{ik}
\partial^2/\partial x^{i}\partial x^{k}$ is the $D$-dimensional
Laplace operator (see \cite{EZ3} for details). We also introduced
the non-relativistic three-dimensional mass density  $\rho_3 =m
\delta({\bf r}_3)$, which is connected with the $D$-dimensional mass
density $\rho_D=\rho_3/V_{D'}$. Here, $V_{D'}$ is the total volume
of the internal spaces. For example, if the i-th torus has periods $a_{(i)j}$,
then $V_{D'}=\prod_{i=1}^N\prod_{j=1}^{d_i} a_{(i)j}$.
%and in the definition of $\mu_i$ we should keep in mind that $d_0=3$.

Now, we want to define the components of the energy-momentum tensor
with the help of Einstein equation
%%%%%%
%%%%
\be{3.7} R_{ik}=\frac{2S_D\tilde G_{\mathcal{D}}}{c^4}\left(T_{ik}
-\frac{1}{D-1}g_{ik}T\right)\, ,
\ee
%%%%
where $S_D=2\pi^{D/2}/\Gamma (D/2)$ is the total solid angle
(surface area of the $(D-1)$-dimensional sphere of unit radius)
and $\tilde G_{\mathcal{D}}$ is the gravitational constant in the
$(\mathcal{D}=D+1)$-dimensional spacetime. Introducing the quantity
$k_{\mathcal{D}}\equiv 2S_D\tilde G_{\mathcal{D}}/c^4$ and keeping in
mind that we are considering compact astrophysical objects at rest in our
three-dimensional space (what results in $T_{11}=T_{22}=T_{33}=0$),
we arrive at the following Einstein equations:
%%%%%%
\ba{3.8}
&{}&\frac{1}{2}k_N\rho_3 c^2\approx k_{\mathcal{D}}\left(T_{00}-
\frac{1}{D-1}Tg_{00}\right)\, ,\\
&{}&\frac{1}{2}(1-\tau)k_N\rho_3 c^2\approx k_{\mathcal{D}}\left(-\frac{1}{D-1}
Tg_{\alpha\alpha}\right)\, ,\label{3.9}\\
&{}&\frac{1}{2}\gamma_{i}k_N\rho_3 c^2\approx k_{\mathcal{D}}\left(T_{\mu_i\mu_i}
-\frac{1}{D-1}Tg_{\mu_i\mu_i}\right)\, .\label{3.10}
\ea
%%%%%%

Therefore, the required components of the energy-momentum tensor are

%%%%%%
\ba{3.11}
T_{00}&\approx&\frac{k_N V_{D'}}{k_{\mathcal{D}}}\left(1
-\frac{\tau}{2}\right)\rho_D c^2,\quad T_{\alpha\alpha}=0\, ,\\
T_{\mu_i\mu_i}&\approx&\frac{k_N V_{D'}(\gamma_{i}-1+\tau)}{2k_{\mathcal{D}}}
\rho_D c^2 \label{3.12}\, .
\ea
%%%%%%
The equation for the 00-component shows that the parameter $\tau$ cannot be
equal to 2 because for $\tau =2$ we get $T_{00}=0$, what corresponds to the
uninteresting case of absence of matter. Moreover, $T^0_0=\varepsilon $ is the
energy density of matter. Therefore, up to the terms of the order of $1/c^2$, we have $T_{00}\approx
\varepsilon \approx \rho_D c^2$. This requires the following relation between
Newtonian and multidimensional gravitational constants \cite{footnote1b}:
%%%%%
\be{3.13}
k_N =\frac{2}{2-\tau}\kappa_{\mathcal{D}}/V_{D'} \; \Longrightarrow \;
4\pi G_N = \frac{2}{2-\tau}S_D \tilde G_{\mathcal{D}}/V_{D'}\, .
\ee
%%%%
In the particular case of a point-like massive source, this relation was given
in \cite{EZ4,EZ2}. From Eqs. \rf{3.11} and \rf{3.12} we also obtain the relation
%%%%
\be{3.14}
T_{\mu_i\mu_i}\approx \frac{\gamma_{i}-1+\tau}{2-\tau} T_{00}\, .
\ee
%%%%%%
Taking into account that, up to the terms of the order of $1/c^2$, components $T_{\mu_i\mu_i}$ define
pressure in the i-th internal space ($T_{\mu_i\mu_i}\approx p_i$), we get from Eq.
\rf{3.14} the following equations of state in these spaces:
%%%%%%
\be{3.15}
p_i = \frac{\gamma_{i}-1+\tau}{2-\tau}\, \varepsilon \, ,\quad  i=1,\ldots ,N.
\ee
%%%%%%%
Since $T_{11}=T_{22}=T_{33}=0$, in our three-dimensional space we have dust-like
equation of state, namely: $p_0=0$. In the case of a point-like mass, the parameters
$\gamma_i$ satisfy the condition \rf{2.8}. It can be easily seen that for these
values of $\gamma_i$, all $T_{\mu_i\mu_i}$ are equal to zero. Therefore, in this case,
$T_{00}$ is the only non-zero component in the external space, as well as in all
internal spaces, and we have the same dust-like equations of state in all spaces, namely: $p_i=0\, ,\;
i=0,\ldots ,N$.

%%%%%%%%%%%%%%%%%%%%%%%%%%%%%%%%%%%%%%%%%%%%%%%%%%%%%%%%%%%%%%%%%%%%%%%%%%%%%%%%%%%%%%%%%%%%%%%%
%%%%%%%%%%%%%%%%%%%%%%%%%%%%%%%%%%%%%%%%%%%%%%%%%%%%%%%%%%%%%%%%%%%%%%%%%%%%%%%%%%%%%%%%%%%%%%%%

\section{\label{sec:4}Experimental restrictions on solitons. Latent solitons}

In this section, we want to get the experimental restrictions for the parameters
$\gamma_i$. This can be done with the help of the parameterized post-Newtonian (PPN)
formalism. According to the PPN formalism (see, e.g., \cite{Will,Straumann}), the
four-dimensional static spherically symmetric line element in isotropic coordinates
is parameterized as follows:
%%%%%%
\be{4.1}
ds^2=\left(1-\frac{r_g}{r_3}+\beta\frac{r_g^2}{2r_3^2} \right)c^2dt^2 -
\left(1+\gamma \frac{r_g}{r_3}\right)\sum_{i=1}^{3}\left(dx^i\right)^2 \, .
\ee
%%%%%%
In general relativity we have $\beta=\gamma=1$. To get $\beta$ and $\gamma$ in
the case of the soliton solution \rf{2.1}, it is sufficient  to compare the metric
coefficients in Eq. \rf{4.1} with the corresponding asymptotic expressions \rf{3.1}
and \rf{3.2}, what immediately gives the soliton PPN parameters
%%%%%%
\be{4.2}
\beta_s =1\, , \quad \gamma_s=1-\tau\, .
\ee
%%%%%%
With the help of these PPN parameters, we can easily get formulas for the famous
gravitational experiments \cite{EZ3,Will,Will2}:

%\vspace{0.4cm}

(i) {\it Perihelion shift}
%%%%%%
\ba{4.3}
&{}&\delta\psi=\frac{6\pi mG_N}{\lambda \left(1-e^2\right)c^2}\frac{1}{3}
(2+2\gamma_s-\beta_s)\nn \\
&=&\frac{6\pi mG_N}{\lambda \left(1-e^2\right)c^2}\frac{3-2\tau}{3}=
\frac{\pi r_g}{\lambda \left(1-e^2\right)}(3-2\tau)\, , \ea
%%%%%
where $\lambda$ is the semi-major axis of the ellipse and $e$ is its
eccentricity.

%\vspace{0.4cm}

(ii) {\it Deflection of light}
%%%%%
\be{4.4} \delta\psi=(1+\gamma_{s})\frac{r_g}{\rho}=(2-\tau)\frac{r_g}{\rho}\, , \ee
%%%%%
where $\rho$ is the distance of closest approach (impact parameter) of the ray's
path to the gravitating mass $m$.

%\vspace{0.3cm}
(iii) {\it Time delay of radar echoes (Shapiro time-delay effect)}
%%%%%%
\ba{4.5}
&{}&\delta t=(1+\gamma_{s})\frac{r_g}{c}\ln\left(\frac{4r_{Earth}r_{planet}}
{R_{Sun}^2}\right)\nn \\
&=& (2-\tau)\frac{r_g}{c}\ln\left(\frac{4r_{Earth}r_{planet}}{R_{Sun}^2}\right)\, . \ea
%%%%%%%%

Comparison of the formulas \rf{4.3}-\rf{4.5} with experimental data gives the
possibility to restrict parameters of the soliton solutions. In fact, we can also get
it directly from experimental restriction on the PPN parameter $\gamma$. The tightest
constraint on $\gamma$ comes from  the Shapiro time-delay experiment using the
Cassini spacecraft, namely:
$\gamma-1 =(2.1\pm 2.3)\times 10^{-5}$ \cite{Will2,JKh,Bertotti}. Thus, from
Eq. \rf{4.2} we find that the solitonic parameter $\tau$ should satisfy
the condition~\cite{footnote2}
%%%%%%
\be{4.6}
\tau = -(2.1\pm 2.3)\times 10^{-5}\, .
\ee
%%%%%%

In the case of the point-like massive soliton described by Eqs. \rf{2.7}-\rf{2.9}, we have
$\tau =D'/(1+D')\sim \mathcal{O}(1)$, what obviously contradicts Eq. \rf{4.6}
(in accordance with the results of \cite{EZ4,EZ3}).

Equation \rf{4.2} shows that there is a very interesting class of solitons which
are defined by the condition
%%%%%%
\be{4.7}
\tau =\sum_{i=1}^N d_i \gamma_i = 0\, .
\ee
%%%%%%
Counting only with the gravitational experiments mentioned above, it is impossible
to differ these Kaluza-Klein solitons from general relativity because they have
$\gamma_s=1$ as in general relativity~\cite{footnote3}.
For this reason, we call these solutions {\it{latent solitons}}. For these latent solitons,
equations of state \rf{3.15} in the internal spaces are reduced to
%%%%%%
\be{4.8}
p_i = \frac{\gamma_{i}-1}{2}\, \varepsilon \, ,\quad  i=1,\ldots ,N.
\ee
%%%%%%%
Black strings ($N=1,\, d_1=1$) and black branes ($N>1$) are characterized by the
condition that all $\gamma_i =0,\, i\geq 1$. Obviously, they belong to the class of
latent solitons and they have the equations of state
%%%%%%
\be{4.9}
p_i = -\frac{1}{2}\, \varepsilon \, ,\quad  i=1,\ldots ,N.
\ee
%%%%%%%
It is known (see, e.g., \cite{EZ4,Zhuk}) that in the case of three-dimensional external
space such equations of state are the only ones which do not spoil the condition of the
internal space stabilization for the compact astrophysical objects with the dust-like
equation of state $p_0=0$ (in the external space).
%In the next section we show that only these equations of state do not spoil the condition of the internal space stabilization.
Therefore, non-zero parameters $\gamma_i$ can be treated as a measure of the latent
soliton destabilization \cite{footnote4}.

We would like to stress the following: It is well known that black strings/branes
have the topology (4-dimensional Schwarzschild spacetime) $\times$ (flat internal
spaces). In this case, it does not seem surprising that gravitational experiments
lead to the same results as for general relativity. However, the latent solitons, in the
general case, do not have either Schwarzschildian metrics for 4-dimensional part of spacetime,
nor flat metrics for the extra dimensions. Nevertheless, within the considered
accuracy, it is also impossible to distinguish them from general relativity.
This is really surprising.

To conclude this section, we would like to mention that the relation between
Newtonian and multidimensional gravitational constants for latent solitons is
reduced to the following equation:
%%%%%
\be{4.10}
4\pi G_N = S_D \tilde G_{\mathcal{D}}/V_{D'}\, .
\ee
%%%%%%%

%%%%%%%%%%%%%%%%%%%%%%%%%%%%%%%%%%%%%%%%%%%%%%%%%%%%%%%%%%%%%%%%%%%%%%%%%%%%%%%%%%%%%%%%%%%%%%%%
%%%%%%%%%%%%%%%%%%%%%%%%%%%%%%%%%%%%%%%%%%%%%%%%%%%%%%%%%%%%%%%%%%%%%%%%%%%%%%%%%%%%%%%%%%%%%%%%

\section{\label{sec:5}Experimental restrictions on the equations of
state of a multidimensional perfect fluid}

In this section, we want to show that for static spherically symmetric
perfect fluid with dust-like equation of state in the external space, the
condition $h_{00}=h_{\alpha\alpha}$ (which provides the agreement with the
gravitational experiments at the same level of accuracy as general relativity)
results in the latent soliton condition \rf{4.7}, and equations of state \rf{4.8},
together with the condition $R_{\mu_i\mu_i}=0 \Longrightarrow  h_{\mu_i\mu_i}=0$,
leads to the stability condition \rf{4.9} and singles out $d_0=3$ for the
number of the external dimensions.

Let us consider a static spherically symmetric perfect fluid with energy-momentum
tensor
%%%%%%%
\ba{5.1}  {T^i}_k &=& {\rm diag\ } (\, \varepsilon, \underbrace{-p_0,\ldots ,
-p_0}_{\mbox{$d_0$ times} }, \; \, \nn\\
&{}&\underbrace{-p_1,\ldots ,-p_1}_{ \mbox{$d_1$ times}}, \ldots ,
\underbrace{-p_N,\ldots ,-p_N}_{ \mbox{$d_N$ times}}\, )\, . \ea
%%%%%%%%%
We recall that we are using the notations: $i,k = 0,1,\ldots,D\, ; \; a,b = 1,
\ldots ,D\, ; \; \alpha,\beta =1,\ldots,d_0$ and $\mu_i = 1+ \sum_{j=0}^{i-1} d_j,
\ldots, d_i + \sum_{j=0}^{i-1} d_j \, ,\quad i=1,\ldots,N$. For static spherically
symmetric configurations we have $g_{0a}=0$ and $g_{ab}=0\, ,\,  a\neq b$. Since we
want to apply this model to ordinary astrophysical objects, where the condition
$T^0_{\;\;0}\gg|T^\alpha_{\;\;\alpha}|$ usually holds, we assume the dust-like
equation of state in the $d_0$-dimensional external space, namely $p_0=0$, but we leave
equations of state arbitrary in the i-th internal space, namely $p_i=\omega_i
\varepsilon$. Obviously, $\varepsilon$ is equal to zero outside the compact
astrophysical objects. Moreover, we consider the weak-field approximation, in which the metric
coefficients can be expressed in the form
%%%%%%
\be{5.2}
g_{00}\approx 1+h_{00}\, ,\quad g_{aa}\approx -1+h_{aa}\, , \quad h_{00},h_{aa}
\sim O\left(1/c^2\right)\, .
\ee
%%%%%%
As an additional requirement, we impose that the considered configuration does
not contradict the observations. It will be so if the following conditions hold:
$h_{00}=h_{\alpha\alpha}$ and $h_{\mu_i\mu_i}=0$ (see Ref. \cite{EZ4}). In what
follows, we define which equations of state are obtained as a result of these restrictions.

Taking into account that $T=\sum_{i=0}^D {T^i}_i=\varepsilon
(1-\sum_{i=1}^N\omega_i d_i)\, ,\; T_{\alpha\alpha}=0\, ,\; \varepsilon \sim O(c^2)$,
and, up to terms of the order of $c^2$, that $T_{00}\approx {T^0}_0\, ,\; T_{\mu_i\mu_i}
\approx -{T^{\mu_i}}_{\mu_i}$, we get from the Einstein equation \rf{3.7} the
non-zero components of Ricci tensor (up to the order of $1/c^2$):
%%%%%%
\ba{5.3} R_{00}&\approx& \frac{\varepsilon k_{\mathcal{D}}}{D-1}\;
\left[d_0-2+\sum_{i=1}^{N}d_{i}(1+\omega_{i})\right]\, ,\\
R_{\alpha\alpha} &\approx& \frac{\varepsilon k_{\mathcal{D}}}{D-1}\;
\left[1-\sum_{i=1}^{N}d_{i}\omega_{i}\right]\, ,\label{5.4}\\
R_{\mu_i\mu_i} &\approx& \frac{\varepsilon k_{\mathcal{D}}}{D-1}\times\nn\\
&\times& \left[\omega_i\left(\sum_{j=0}^{N}{'}d_{j}-1\right)+1-
\sum_{j=1}^{N}{'}d_{j}\omega_{j}\right] \label{5.5}\, , \ea
%%%%%%
where $k_{\mathcal{D}}\sim O(1/c^4)$ is defined in Sec. III, and the prime in the
summation of Eq. \rf{5.5} means that we must not take into account the i-th term.
Equations \rf{5.3} and \rf{5.4} show that the $R_{00}$ and $R_{\alpha\alpha}$ components
are related as follows:
%%%%%
\be{5.6}
R_{\alpha\alpha} = \frac{1-\sum_{i=1}^{N}d_{i}\omega_{i}}{d_0 - 2 +
\sum_{i=1}^{N}d_{i}(1+\omega_i)}R_{00}\, .
\ee
%%%%%%
On the other hand, in the weak-field limit the components of Ricci tensor read
%%%%%%%
\be{5.7} R_{00}\approx \frac12 \triangle h_{00}\, ,\quad  R_{aa}\approx \frac12
\triangle h_{aa}\, ,\quad  a=1,\ldots ,D\, , \ee
%%%%%%%
where as usual we can put $h_{00}\equiv 2\varphi/c^2$, and $\triangle$ is
$D$-dimensional Laplace operator defined in Eqs. \rf{3.4}-\rf{3.6}. Therefore, from Eqs.
\rf{5.6} and \rf{5.7} we obtain
%%%%%%%
\be{5.8} h_{\alpha\alpha}=\frac{1-\sum_{i=1}^{N}d_{i}\omega_{i}}{d_0 - 2 +
\sum_{i=1}^{N}d_{i}(1+\omega_i)}h_{00}\, ,\quad \alpha =1,\ldots ,d_0\, . \ee
%%%%%%
As we have mentioned above, to be in agreement with experiments we should demand
$h_{\alpha\alpha}=h_{00}$, what leads to the following restriction on the parameters
$\omega_i$ of the equations of state:
%%%%%
\be{5.9}
3-d_0 - \sum_{i=1}^{N}d_{i} = 2\sum_{i=1}^{N}d_{i}\omega_{i}\, .
\ee
%%%%%
In the case of three-dimensional external space ($d_0=3$), this constraint is reduced to
%%%%
\be{5.10} \sum_{i=1}^N d_i\left(\omega_i+\frac12\right)=0\, . \ee
%%%%%
If we parameterize
%%%%%
\be{5.11}
\omega_i = \frac{\gamma_i-1}{2}\, ,\quad i=1,\ldots ,N\, ,
\ee
%%%%%%
then we arrive at the latent soliton condition \rf{4.7}. Therefore, {\em the
demand that multidimensional perfect fluid with dust-like equation of state ($p_0=0$)
in the external space provides the same results for gravitational experiments as general
relativity, leads to the latent soliton equations of state \rf{4.8} in the
internal spaces}. However, it is known (see Refs. \cite{EZ4,Zhuk}) that the internal
spaces can be stabilized if multidimensional perfect fluid with $p_0=0$ has the
same equations of state $\omega_i=-1/2$ in all internal spaces and the external space
is three-dimensional ($d_0=3$). In other words, it takes place if all $\gamma_i=0$ in
Eq. \rf{5.11}. Let us show that the additional requirement $R_{\mu_i\mu_i}=0$ ensures the
fulfillment of these conditions. Indeed, from Eq. \rf{5.5}  we get
%%%%%
\be{5.12}
R_{\mu_i\mu_i}=0\quad \Longrightarrow \quad \omega_i =-\frac12\, , \quad i=1,\ldots ,N\, ,
\ee
%%%%%
where we used the constraint \rf{5.9} \cite{footnote5}. Now, the
substitution $\omega_i=-1/2$ in Eq. \rf{5.9} singles out $d_0=3$.  Therefore, {\em the demand
of the internal space stabilization leads, for multidimensional perfect fluid
with $p_0=0$, to the black string/brane equations of state \rf{4.9} in the internal
spaces and, additionally, it selects uniquely the number of the external spaces to be
$d_0=3$}.

To conclude the consideration of this perfect fluid, we want to get the metric
coefficients up to $\mathcal{O}(1/c^2)$ [see Eq. \rf{5.2}]. To do so, it is sufficient
to define the function $\varphi\equiv h_{00}c^2/2$. It can be easily seen from Eqs.
\rf{5.3} and \rf{5.7} that this function satisfies the equation
%%%%%
\be{5.13}
\triangle\varphi =\frac{c^2}{2} \triangle h_{00}  \approx c^2 R_{00} \approx
S_D \tilde G_{\mathcal{D}}\rho_D\, ,
\ee
%%%%%%%
where we have used the constraint \rf{5.9} for arbitrary $d_0$ and the relation
$\varepsilon \approx \rho_D c^2$. Therefore, to get the metric coefficients we
need to solve this equation with proper boundary conditions. We want to reduce
this equation to ordinary Poisson equation in three-dimensional external space
($d_0=3$). To do so, we consider the case in which matter is uniformly smeared over
the extra dimensions, then  $\rho_D=\rho_{3}/V_{D'}$ (see Sec. III). In this case
the non-relativistic potential $\varphi$ depends only on our external coordinates
and $\triangle$ is reduced to three-dimensional Laplace operator $\triangle_3$.
Therefore, Eq. \rf{5.13} is reduced to
%%%%%%
\be{5.14}
\triangle_3 \varphi \approx  (S_D \tilde G_{\mathcal{D}}/V_{D'})\rho_3=4\pi G_N \rho_3\, ,
\ee
%%%%%
where we have used the relation \rf{4.10} between Newtonian and multidimensional gravitational
constants. Equation \rf{5.14} is the usual Poisson equation. It is worth noting that $\rho_{3}=0$
outside the compact astrophysical object and it is necessary to solve Eq. \rf{5.14}
inside and outside of the object, and to match these solutions at the boundary.

%%%%%%%%%%%%%%%%%%%%%%%%%%%%%%%%%%%%%%%%%%%%%%%%%%%%%%%%%%%%%%%%%%%%%%%%%%%%%%%%%%%%%%%%%%%%%%%%
%%%%%%%%%%%%%%%%%%%%%%%%%%%%%%%%%%%%%%%%%%%%%%%%%%%%%%%%%%%%%%%%%%%%%%%%%%%%%%%%%%%%%%%%%%%%%%%%
%%%%%%%%%%%%%%%%%%%%%%%%%%%%%%%%%%%%%%%%%%%%%%%%%%%%%%%%%%%%%%%%%%%%%%%%%%%%%%%%%%%

\section{Conclusions}

In the first part of our investigations (Sec. \ref{sec:2}-\ref{sec:4}), we considered
the most general (known to us) soliton solution in Kaluza-Klein models with
toroidal compactification of the extra dimensions. Here, each $d_i$-dimensional torus
has its own scale factor $C_i(r_3),\; i=1,\ldots ,N$, which is characterized by the
parameter $\gamma_i$. A distinctive feature of these solutions is that their metric
coefficients depend only on the length of a radius vector $r_3$ of the external (our)
space \cite{footnote6}. This happens when the matter source is uniformly smeared over the extra dimensions. In this
case, the non-relativistic gravitational potential exactly coincides with the Newtonian
one. Among the soliton solutions, we sorted out one which corresponds to a point-like mass.
This solution is of special interest because it generalizes the well known point-like mass
approach of general relativity, which works very well to describe the known gravitational
experiments. Then, we investigated the weak-field limit and obtained (in the general case)
the equations of state for the soliton matter source. These equations show that in the case
of a point-like mass, $T_{00}$ is the only non-zero component of the energy-momentum tensor, and
the equations of state in the external and internal spaces correspond
to dust ($p_i=0,\; i=0,\dots ,N$). We also used the weak-field limit to get the experimental
restrictions on the parameters of the soliton solutions. To get it, we found the
parameterized post-Newtonian (PPN) parameters $\beta_s$ and $\gamma_s$ for the soliton
solutions. This gave us a possibility to derive formulas for perihelion shift, deflection
of light and time delay of radar echoes. For soliton solutions, the parameter $\beta_s=1$
coincides with the one in general relativity. However, the parameter $\gamma_s =1-\tau=1-
\sum_{i=1}^Nd_i\gamma_i$ is different from its general relativistic value ($\gamma = 1$).
The PPN parameter $\gamma$ is strongly restricted by the Shapiro time-delay
measurements obtained from
the Cassini spacecraft. With the help of this bound, we obtained the limitation on the
soliton parameter, namely: $\tau = -(2.1\pm 2.3)\times 10^{-5}$. The point-like mass soliton
considerably contradicts this restriction. Obviously, solutions with $\tau =
\sum_{i=1}^Nd_i\gamma_i=0$ satisfy this bound. This is a new class of soliton
solutions. For them we have $\gamma_s =1$ as in general relativity, and therefore it is impossible
to differ experimentally these Kaluza-Klein solitons from general relativity. For this reason we
call these solutions {\em latent solitons}. For the latent solitons, the non-relativistic
equations of state are $p_0=0$ in the external (our) space and $p_i=[(\gamma_i-1)/2]\varepsilon$
in the internal spaces. All these results were obtained for the realistic case of
three-dimensional external space. We would like to stress once again that latent
solitons satisfy the gravitational experiments mentioned above at the same level
of accuracy as general relativity. Black strings ($N=1,\, d_1=1$) and black branes
($N>1$) are characterized by the condition that all $\gamma_i =0,\, i\geq 1$. Obviously,
they belong to the class of latent solitons and they have the equations of state $p_i =
-\varepsilon/2 \, ,\quad  i=1,\ldots ,N$. It is known (see Refs. \cite{EZ4,Zhuk})
that in the case of three-dimensional external space such equations of state are the
only ones which do not spoil the condition of the internal space stabilization for the
compact astrophysical objects with the dust-like equation of state ($p_0=0$) in the
external space.

In the second part of our investigations (Sec. \ref{sec:5}), we considered a
multidimensional static spherically symmetric perfect fluid with dust-like equation of state
($p_0=0$) in the external space and arbitrary equations of state ($p_i=\omega_i\varepsilon$)
in the internal spaces. The number of external spatial dimensions $d_0$ was left
arbitrary. In the case $d_0=3$, such perfect fluid describes observable astrophysical
objects (e.g., the Sun) in Kaluza-Klein models. We performed our analysis in the weak-field
limit where the metric coefficients can be expressed in the form $g_{00}\approx 1+h_{00},
\; g_{aa}\approx -1+h_{aa},\; a=1,\ldots ,D$, and $h_{00},h_{aa} \sim
O\left(1/c^2\right)$. (We recall that $h_{\alpha\alpha},\; \alpha=1,\ldots , d_0$, describe
perturbations in the external space and $h_{\mu_i\mu_i}, i=1,\ldots ,N$, describe
perturbations in the internal spaces). We have shown that the demand of agreement
with the gravitational experiments at the same level of accuracy as general relativity,
$h_{00}=h_{\alpha\alpha}$, results in a constraint for the parameters $\omega_i$,
which exactly coincides with the latent soliton condition, $\sum_{i=1}^Nd_i\gamma_i=0$, in
the case $d_0=3$. In other words, for $d_0=3$, the equations of state in the internal spaces
are $p_i=[(\gamma_i-1)/2]\varepsilon$. The additional requirement $h_{\mu_i\mu_i}=0$,
together with the previous condition $h_{00}=h_{\alpha\alpha}$, (i) leads
to the equation $\omega_i=-1/2$ (i.e. $\gamma_i=0$) and (ii) singles
out $d_0=3$ \cite{footnote7}. Therefore, we arrived at the black string/brane equations of
state in the internal spaces. Precisely these equations of state (supplemented by the
condition $p_0=0$) satisfy the necessary condition of the internal space stabilization
in the case $d_0=3$ \cite{EZ4,Zhuk}. We see that the first and second parts of our
investigations agree with each other, as expected.

We can summarize the main conclusion of our paper as follows. For compact
astrophysical objects with dust-like equation of state in the external space
($p_0=0$), the demand of the agreement with the gravitational experiments
requires the condition \rf{4.6}, namely: $\tau = -(2.1\pm 2.3)\times 10^{-5}$.
However, to be at the same level of accuracy as general relativity, we
must have $\tau=0$. In other words, we should consider the latent solitons
with equations of state \rf{4.8} in the internal spaces (in the case
$d_0=3$). Moreover, the condition of stability of the internal spaces
singles out black strings/branes from the latent solitons and leads
uniquely to $p_i=-\varepsilon/2$ as the black string/brane equations of
state in the internal spaces, and to the number of the external dimensions
$d_0=3$.
%As we noted in our previous paper \cite{EZ4},
The main problem with the black strings/branes is to find a physically
reasonable mechanism which can explain how the ordinary particles forming
the astrophysical objects can acquire rather specific equations of state
($p_i=-\varepsilon/2$) in the internal spaces.

%\centerline \mbox{} \\ {\bf ACKNOWLEDGEMENTS}\\

\section*{ACKNOWLEDGEMENTS}

O. M., L. C., and A. Zh. would like to
acknowledge partial financial support from
Coordena\c{c}\~ao de Aperfei\c{c}oamento de Pessoal
de N\'\i vel Superior (CAPES). O. M. and L. C. also acknowledge partial
financial support from Conselho Nacional de Desenvolvimento Cient\'ifico
e Tecnol\'ogio (CNPq).
M. E. and A. Zh. acknowledge partial financial support from
``Cosmomicrophysics" Programme of the Physics and Astronomy
Division of the National Academy of Sciences of Ukraine.
A. Zh. thanks also the Faculdade de F\'\i sica of the
Universidade Federal do Par\'{a} for their kind hospitality
during the preparation of this work.
%%%%%%%%%%%%%%%%%%%%%%%%%%%%%%%%%%%%%%%%%%%%%%%%%%%%%%%%%%%%%%%%%%%%%%%%%%
%%%%%%%%%%%%%%

%%%%%%%%%%%%%%%%%%%%%%%%%%%%%%%%%%%%%%%%%%%%%%%%%%%%%%%%%%%%%%%%%%%%%%%%%%%%%%%%%%
%%%%%%%%%%%%%%%%%%%%%%%%%%%%%%%%%%%%%%%%%%%%%%%%%%%%%%%%%%%%%%%%%%%%%%%%%%%%%%%%%%


\begin{thebibliography}{99}
\bibitem{footnote1}
In the literature,
a different parameterization (in terms of parameters $k$ and $\varepsilon$)
is commonly used (see, e. g., Ref.~\cite{Davidson})
in the case of one internal space $N=1$. The relation between both
of these parameterizations is the following: $\theta =\varepsilon k$ and
$\gamma_{1}\equiv \gamma =1/k$. Then, $\theta \gamma =\varepsilon$ and $\theta
(1-\tau)=\varepsilon (k-d)$, where $d\equiv d_1$.
\bibitem{footnote1a}
A particular soliton solution was also found in Ref.~ \cite{Cavaglia}.  In the case of four external
spacetime coordinates and $D'$ internal spatial coordinates, the parameters in the corresponding solution (25)
in Ref.~ \cite{Cavaglia} are: $q=2,\, p=D'$ and $\Delta^2 = 2(D'+1)/(D'+2)$. If we rewrite this solution
in the isotropic coordinates, then we immediately get the connection with our parameters,
namely: $\gamma_i=-1, \, i=1,\ldots ,N , \, \tau = -D' , \, \sigma = D' $  and $\theta = \sqrt{2/[(D'+1)(D'+2)]}$.
\bibitem{footnote1b}
It is worth noting the singular nature of $\tau=2$, which follows from Eq. \rf{3.11},
as we require the $T_{00}$ component of the energy-momentum tensor to coincide with the energy density
$\rho_D c^2$ in the non-relativistic limit. Such relationship between $T_{00}$ and $\rho_D c^2$ provides
the correct transition to the Newtonian approximation. This physical condition leads to equation \rf{3.13},
which diverges for $\tau = 2$.
\bibitem{footnote2}
In the case of one $d$-dimensional internal space ($N=1$), we can rewrite this
bound with respect to the parameter $k$ defined in~\cite{footnote1} as
$|k|\geq d\times 2.3\times10^4$. This inequality shows that the increase of
the number of internal space dimensions $d$ imposes stronger restrictions on $k$.
\bibitem{footnote3}
It can be easily
seen from Eqs. \rf{3.1}-\rf{3.3} that the parameter $\tau$ defines also the difference
between perturbations $h_{00}$ and $h_{\alpha\alpha}$: $\; h_{00}-h_{\alpha\alpha}=
-\tau r_g/r_3$. Precisely because of this difference, gravitational experiments in
Kaluza-Klein models and in general relativity lead to different results. When $\tau \to 0$,
this difference disappears. The additional limit $\gamma_i \to 0 \Longrightarrow
h_{\mu_i\mu_i}\to 0$ provides stabilization of the internal spaces \cite{EZ4}. It is worth
noting that this instability is different from the
Gregory-Laflamme one for black strings, as it clearly follows from appendix in Ref. \cite{EZ4}.
\bibitem{footnote4}
If we rewrite equations of
state in the form $p_i=(\alpha_i-1)\varepsilon\, , \;  i=0,\ldots ,N $, then for
the latent solitons we have $\alpha_0=1,\, \alpha_i=(1+\gamma_i)/2\, ,\;  i=1,
\ldots ,N$. For these values of $\alpha_0$ and \,$\alpha_i$, we get on the right-hand
side of Eq. (A15) in Ref. \cite{EZ4} the terms $(\gamma_id_i/2)\kappa_N\rho_3$.
These terms are dynamical functions because of the dynamical behavior of the energy
density $\rho_3$. This results in a violation of the necessary condition for the internal
space stabilization.
\bibitem{footnote5}
It can be also easily
seen that $R_{\mu_i\mu_i}=0\; \Longrightarrow \; \triangle h_{\mu_i\mu_i}=0$, what,
together with the boundary conditions (finiteness of $h_{\mu_i\mu_i}$ at $r_3=0$
and $h_{\mu_i\mu_i}\to 0$ for $r_3\to +\infty$), gives $h_{\mu_i\mu_i}=0$.
\bibitem{footnote6}
We assumed the external space to be three-dimensional ($d_0=3$), because we want to
establish the relation between the considered solitons and the observations.
\bibitem{footnote7}
It is worth noting that in Ref. \cite{Randall} there is also a suggestion of
explanation why our world is three-dimensional. The reasoning is
based on the behavior of isotropic multidimensional Friedmann universes filled with
a gas of branes. The authors have arrived at the conclusion that such universe
will naturally come to be dominated by 3-branes and 7-branes. Obviously, our conclusion
about the three-dimensionality of our Universe is based on a completely different (anisotropic)
multidimensional model.
\end{thebibliography}
\end{document}